\documentclass[preprints,article,accept,moreauthors,pdftex]{Definitions/mdpi} 

\firstpage{1} 
\pubvolume{xx}
\issuenum{1}
\articlenumber{5}
\pubyear{2019}
\copyrightyear{2019}
\history{Received: date; Accepted: date; Published: date}
\pdfoutput=1

\Title{Time-Resolved XUV absorption spectroscopy and magnetic circular dichroism at the Ni $M_{2,3}$-edges}

\Author{Marcel Hennes $^{1}$*, Benedikt Rösner $^{2}$, Valentin Chardonnet $^{1}$, Gheorghe S. Chiuzbaian $^{1}$, Renaud Delaunay $^{1}$, Florian Döring $^{2}$, Vitaliy A. Guzenko $^{2}$, Michel Hehn $^{3}$, Romain Jarrier $^{1}$, Armin Kleibert $^{2}$, Maxime Lebugle $^{2}$, Jan Lüning $^{1}$, Gregory Malinowski $^{3}$, Aladine Merhe $^{1}$, Denys Naumenko $^{4}$, Ivaylo P. Nikolov $^{4}$, Ignacio Lopez-Quintas $^{4}$, Emanuele Pedersoli $^{4}$, Tatiana Savchenko $^{2}$, Benjamin Watts $^{2}$, Marco Zangrando $^{4}$,  Christian David $^{2}$, Flavio Capotondi $^{4}$, Boris Vodungbo $^{1}$ and Emmanuelle Jal $^{1,}$*}

\AuthorNames{Marcel Hennes, Benedikt Rösner,Valentin Chardonnet, Gheorghe S. Chiuzbaian, Renaud Delaunay, Florian Döring, Vitaliy A. Guzenko, Michel Hehn, Romain Jarrier, Armin Kleibert, Maxime Lebugle, Jan Lüning, Aladine Merhe, Denys Naumenko, Ivaylo P. Nikolov, Ignacio Lopez-Quintas, Emanuele Pedersoli, Tatiana Savchenko, Benjamin Watts, Marco Zangrando,  Christian David, Flavio Capotondi, Boris Vodungbo and Emmanuelle Jal}

\address{%
$^{1}$ \quad Sorbonne Université, CNRS, Laboratoire de Chimie Physique – Matière et Rayonnement, LCPMR, Paris 75005, France\\
$^{2}$ \quad Paul Scherrer Institut, Villigen PSI 5232, Switzerland \\
$^{3}$ \quad Institut Jean Lamour, Université de Lorraine, Nancy 54011, France \\
$^{4}$ \quad FERMI, Elettra-Sincrotrone Trieste, 34149 Basovizza, Trieste, Italy}
\corres{Correspondence: marcel.hennes@sorbonne-universite.fr; emmanuelle.jal@sorbonne-universite.fr}

\abstract{Ultrashort optical pulses can trigger a variety of non-equilibrium processes in magnetic thin films affecting electrons and spins on femtosecond timescales. In order to probe the charge and magnetic degrees of freedom simultaneously, we developed an x-ray streaking technique that has the advantage of providing a jitter-free picture of absorption cross section changes. In this paper, we present an experiment based on this approach which we performed using five photon probing energies at the Ni $M_{2,3}$-edges. This allowed us to retrieve the absorption and magnetic circular dichroism time traces, yielding detailed information on transient modifications of electron and spin populations close to the Fermi level. Our findings suggest that the observed charge and magnetic dynamics both depend on the XUV probing wavelength, and can be described, at least qualitatively, by assuming ultrafast energy shifts of the electronic and magnetic elemental absorption resonances, as reported in recent work. However, our analysis also hints at more complex changes, highlighting the need for further experimental and theoretical analysis in order to gain a thorough understanding of the interplay of electronic and spin degrees of freedom in optically excited magnetic thin films.}
% (list three to ten pertinent keywords specific to the article, yet reasonably common within the subject discipline.)}

\begin{document}
%%%%%%%%%%%%%%%%%%%%%%%%%%%%%%%%%%%%%%%%%%

\section{Introduction}
Since its discovery by Beaurepaire and coworkers in 1996 \cite{beaurepaire1996ultrafast}, the phenomenon of laser induced ultrafast demagnetization has attracted world-wide attention and initiated an entirely new research field \cite{kirilyuk2010ultrafast}. However, despite more than 20 years of ongoing experimental and theoretical work, the mechanisms underlying the femtosecond magnetization dynamics in ferromagnetic films following ultrafast optical excitation remain intensively debated \cite{koopmans2010explaining, battiato2010superdiffusive,bierbrauer2017ultrafast,ferte2017element,jal2017structural}. 
The problem is interesting from a fundamental perspective -- to date, it is not clear how angular momentum is transferred between the electron/spin system and the crystalline lattice on sub-picosecond time scales -- but it is also technologically relevant, as using light to steer magnetization on sub-ps timescales might pave the way to ultrafast all-optical spintronics and data storage \cite{stanciu2007AOS,choi2014spin,el-ghazaly2020progress}. 

X-ray experiments are an ideal tool to study the dynamics of the spin, electron and phonon sub-systems, since they combine high spatial resolution and element selectivity. In general, time-resolved scattering and absorption experiments mainly focus on the spin and lattice dynamics, yielding information about superdiffusive spin currents \cite{vodungbo2012laser, pfau2012ultrafast, graves2013nanoscale,weder2020transient}, the role of spin-orbit coupling \cite{boeglin2010distinguishing}, the interaction of different magnetic sublattices \cite{radu2011transient}, and the link between lattice and spin dynamics \cite{henighan2016generation, reid2018beyond, dornes2019ultrafast}. So far, only few  studies have simultaneously looked at electron and spin dynamics \cite{stamm2007femtosecond,kachel2009transient,carva2009influence,boeglin2010distinguishing}. Indeed, performing pure absorption measurements at x-ray free electron lasers (FEL) has been challenging which is due to intrinsic fluctuations of the photon parameters and the lack of a reliable and linear incident intensity monitor \cite{higley2016femtosecond}. However, motivated by time-resolved absorption experiments performed at femtoslicing facilities 
\cite{stamm2007femtosecond,boeglin2010distinguishing}, new technical advances were made recently to probe x-ray absorption dynamics using FEL \cite{higley2016femtosecond, buzzi2017single-shot,jal2019single,rosner2020simultaneous,stamm2020x-ray} as well as high harmonic generation (HHG) sources \cite{willems2015probing,dewhurst2020element}, thereby laying the ground for a systematic analysis of charge and spin dynamics \cite{yao2020distinct,rosner2020simultaneous}.

Within this context, the investigation of the impact of the probing photon energy on the observed electron and magnetization dynamics has attracted increasing interest  \cite{frietsch2015disparate,gort2018early,Somnath2020analysis}. In a recent study, Yao \textit{et al.} \cite{yao2020distinct} provided experimental and theoretical evidence for an electronic and magnetic elemental absorption resonance red shift on ultrafast timescales. Unfortunately, the use of HHG setup in this study \cite{yao2020distinct} results in photon energies separated by approximately 3 eV, which impedes a detailed investigation of the absorption edge itself. To circumvent this shortcoming, we employed x-ray streaking at the XUV-FEL FERMI to probe five energies at the Ni $M_{2,3}$-edges (spanning from 64 to 68 eV). Our results confirm that absorption and magnetic circular dichroism (MCD) spectra are shifting but also hint at a more complicated scenario involving a change of spectra shapes after optical excitation. This calls for additional systematic experimental studies and in-depth theoretical calculations. 

%%%%%%%%%%%%%%%%%%%%%%%%%%%%%%%%%%%%%%%%%%
\section{Experiment}

\begin{figure}[H]
    \centering
    \includegraphics[width=9 cm]{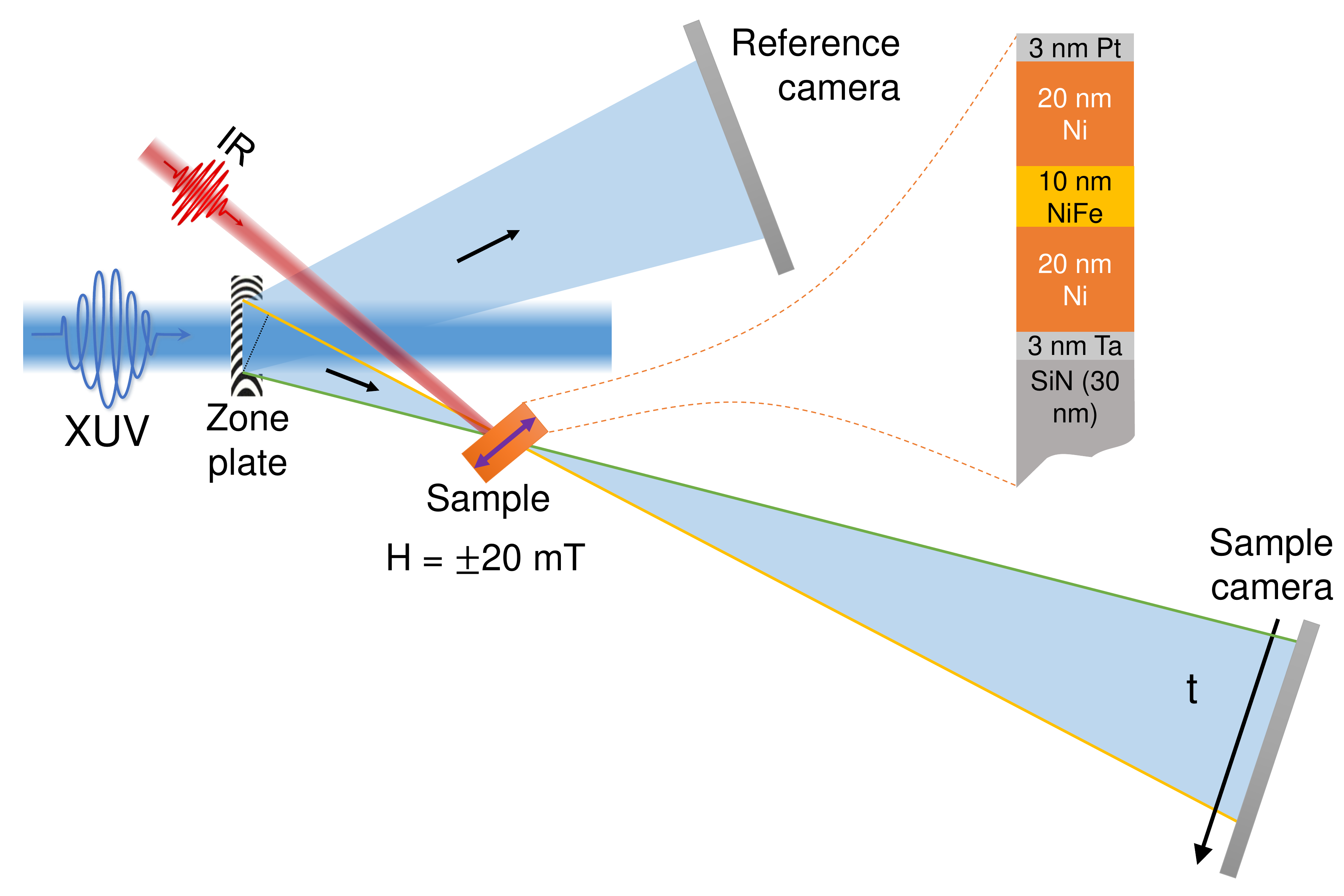}  
    \caption{Experimental setup and sample composition. A circularly polarized XUV beam is diffracted using a zone plate. The positive, first diffraction order is focused on an in-plane magnetized sample, which is pumped using linear infrared pulses. The \textit{sample camera} records the intensity transmitted through the sample $I$, resulting in a spatial encoding of  the time traces. The \textit{reference camera} records the incident intensity $I_0$.}
    \label{Fig0}
\end{figure}

The sample used in our experiment is a 50\,nm thick Ni(20\,nm)/Fe$_{50}$Ni$_{50}$(10nm)/Ni(20\,nm) trilayer\footnote{The presence of the Fe$_{50}$Ni$_{50}$ layer was required for a different femtomagnetism study but is not relevant for the outcome of this study.} deposited via magnetron sputtering on top of Ta(3\,nm)-covered Si$_3$N$_4$(30\,nm) membranes.
%, covered with a 3 nm Ta seed layer. 
A final Pt(3\,nm) capping was used to prevent oxidation (Figure \ref{Fig0}). The magnetic layers are polycrystalline and exhibit an in-plane anisotropy with square magnetic hysteresis loop and a coercive field of 8 mT.

\begin{figure}[H]
    \centering
    \includegraphics[width=14 cm]{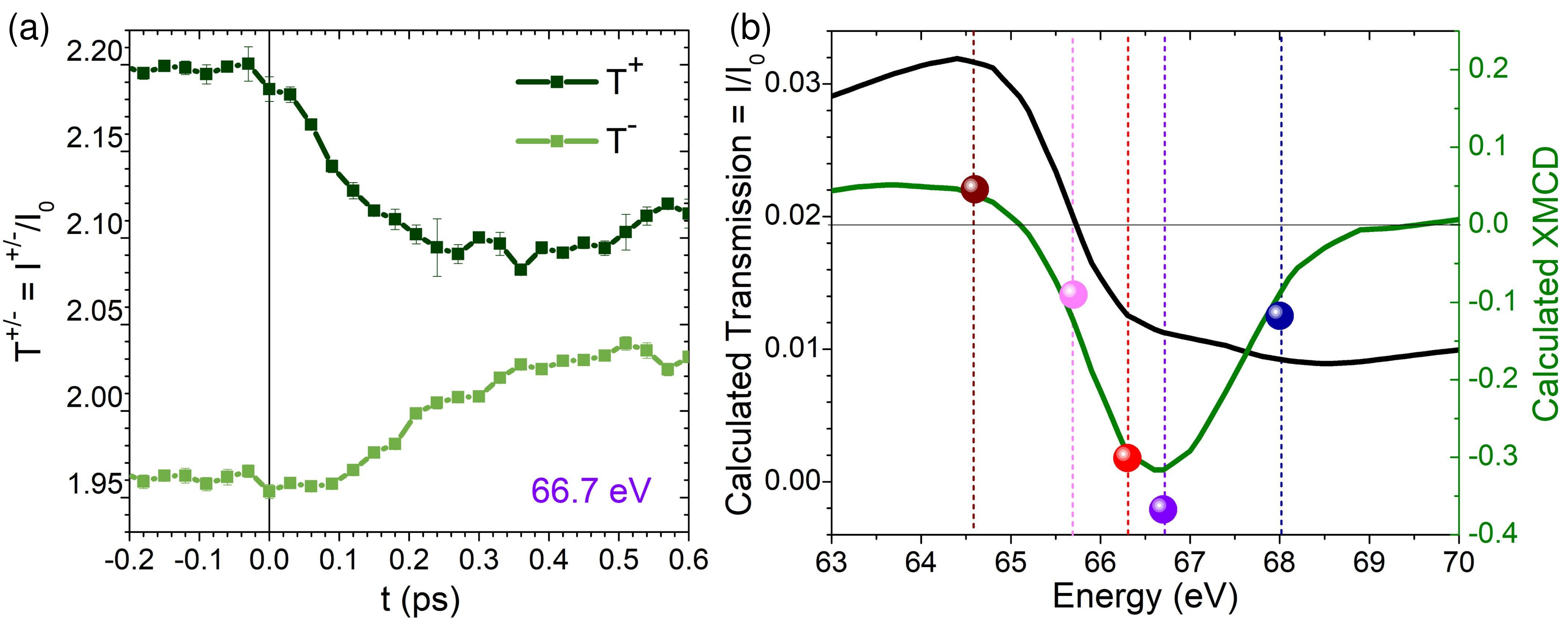}  
    \caption{Characteristic pump-induced changes of $T^+$ and $T^-$ on fs timescales and static energy-dependent transmission and MCD. (\textbf{a}) Transmitted intensity time traces measured for both applied field directions obtained for $E = 66.7$\,eV. (\textbf{b}) Calculated transmission (black) and MCD (green) compared with experimental static MCD (colored dots). }
    \label{Fig1}
\end{figure}

The experiments, based on our recently developed x-ray streaking technique \cite{jal2019single}, were performed at the DiProI end-station \cite{capotondi2013invited} of the seeded XUV-FEL FERMI at Elettra, Trieste \cite{allaria2012highly}. The setup employed in our study is detailed in Figure \ref{Fig0}. As described previously \cite{jal2019single}, our approach is based on the use of a nearly collimated XUV beam with circular polarization, which illuminates an off-axis zone-plate. The converging first order emanating from the zone-plate passes through the sample and its intensity distribution is collected by a CCD camera (named \textit{sample camera}). As shown in Figure \ref{Fig0} the photons detected at the bottom of this camera (yellow line) have travelled a longer path than the ones arriving at the top (green line). This results in a  2.6 ps time difference at the zone plate focus, which becomes spatially encoded in the \textit{sample camera}. To measure the shot to shot variations of the incident x-ray pulses, a second CCD detector (\textit{reference camera}) is used to record the diverging first negative order. By dividing the \textit{sample camera} with the \textit{reference camera} picture, we obtain a measure for the transmission of the sample $T = I/I_0$, where $I$ is the photon intensity after, and $I_0$ the one before the sample. Angular integration of the intensity ratio pictures \cite{jal2019single} eventually yields the transmission time traces shown in Figure \ref{Fig1} (a).

In order to be sensitive to the in-plane magnetization of the Ni layer, the sample was tilted by 18° with respect to the incident beam. To ensure saturation of the magnetization for each pump-probe measurement, an in-plane magnetic field of $\pm$ 20 mT was applied by using an electromagnet. The sample was excited by an 800 nm infrared (IR) laser delivering 60 fs pulses using an incident fluence of 5 mJ/cm$^2$, and probed by circularly polarized 70 fs XUV pulses, with an intensity at the sample plane well below the excitation regime \cite{wang2012femtosecond}. In the sample plane, the size of the IR spot was $\simeq$ 250 $\times$ 165 $\mu$m$^2$ (FWHM), significantly larger than the XUV spot size $\simeq$ 130 $\times$ 70 $\mu$m$^2$ (FWHM). To probe the absorption and MCD around the Ni $M_{2,3}$ resonances, we used five probing XUV wavelengths with energies 64.6 eV, 65.7 eV, 66.3 eV, 66.7 eV and 68 eV. For each energy, we recorded CCD images by accumulation of 150 to 250 shots with and without IR pump for both applied magnetic field directions and both XUV helicities. For symmetry reasons, reversing the XUV probe beam helicity is equivalent to a reversal of the applied magnetic field direction. This allows us to average the measured curves, resulting in two transmitted intensities $T^{\pm} = \frac{I^{\pm}}{I_0}$. We grouped the data such as to obtain a data point every 30 fs, the error bars presented in all our figures reflect this grouping and averaging procedure.

Typical transmitted intensity time traces are plotted in Figure \ref{Fig1} (a) for an XUV energy of 66.7 eV and both external field directions: $H^+$ (dark green) and $H^-$ (light green). As expected, the system is excited on sub-ps timescales: the difference between $T^+$ and $T^-$ is quickly decreasing after $t = 0$, which reflects the strong transient modifications of the magnetic properties of the system. 

To quantify these changes, $T^{\pm}$ data can be used to calculate the absorption $\mu$ and the $MCD$, which are related to the electronic occupation and to the spin state of our magnetic layer, respectively. We therefore use $\mu^{\pm} \propto -\ln{T^{\pm}}$, $\mu = \frac{\mu^+ + \mu^-}{2} \propto \frac{\ln({T^+ \cdot T^-)}}{2}$  and $MCD= \mu^+ - \mu^- \propto \ln{\frac{T^-}{T^+}}$ \cite{stohr2006magnetism}. Quantities extracted from pumped data will be noted $T^{\pm}(t)$, $\mu^{\pm}(t)$, $\mu(t)$ and $MCD(t)$ while static data will be labeled $T^{\pm}_0$, $\mu^{\pm}_0$, $\mu_0$ and $MCD_0$. In order to compare data measured using different XUV energies, we need to look at the relative changes $\frac{\Delta\mu^{\pm}(t)}{\mu^{\pm}_0}$, $\frac{\Delta\mu(t)}{\mu_0}$ and $\frac{\Delta MCD(t)}{MCD_0}$. Here, the prefix $\Delta$ is used as a short notation for a subtraction of the static signal, e.g. $\Delta\mu(t) = \mu(t)-\mu_0$. 

Due to different IR filters used in front of the cameras, experimental transmission data are only known up to a prefactor. This can be seen in Figure \ref{Fig1} (a), where the transmission is superior to 1, which is of course at odds with the fact that the beam intensity is weakened \textit{via} absorption. To correct this artifact, we use the calculated transmission $T_{\mathrm{th}}$ shown in Figure \ref{Fig1} (b) (dark curve) to rescale our data and obtain the proper relative absorption. The calculated transmission and MCD curves presented in Figure \ref{Fig1} (b) have been obtained using tabulated values (CXRO website \cite{cxro}) for Si$_3$N$_4$, Ta and Pt which were combined with literature data for Ni and Fe \cite{willems2019magneto}. Comparison of the calculated MCD signal (green curve in Figure \ref{Fig1} (b)) with our static experimental $MCD_0$ (colored dots) yields good agreement for each energy scrutinized in the present study (Figure \ref{Fig1} (b))\footnote{Note that the experimental MCD spectra does not need to be rescaled since the constant pre-factors cancel out as shown in Equation \ref{eq1}.}. The following equations summarize our approach, and all data presented in the remainder of this work are based on these calculations:
\begin{equation}
    \begin{aligned}
        \frac{\Delta\mu^{\pm}(t)}{\mu^{\pm}_0} &= \frac{\ln{\left(\frac{T^\pm_0}{T^\pm(t)}\right)}}{\ln{\left(\alpha \cdot T^\pm_0\right) }} &  \\
         &  & \\
        \frac{\Delta\mu(t)}{\mu_0} &= \frac{\ln{\left(\frac{T^+_0 \cdot T^-_0}{T^+(t) \cdot T^- (t)}\right)}}{\ln{\left(\alpha^2 \cdot T^+_0 \cdot T^-_0 \right)}} & \text{ with\,\, } \alpha = \frac{2 T_{\mathrm{th}}}{T^+_0 + T^-_0}\\ 
         &  & \\
        \frac{\Delta MCD(t)}{MCD_0} &= \frac{\ln{\left(\frac{T^+_0 \cdot T^-(t)}{T^+(t) \cdot T^-_0}\right)}}{\ln{\left(\frac{T^-_0}{T^+_0}\right)}} & 
    \end{aligned}
    \label{eq1}
\end{equation}

%%%%%%%%%%%%%%%%%%%%%%%%%%%%%%%%%%%%%%%%%%
\section{Results}

\begin{figure}[H]
    \centering
    \includegraphics[width=6.3cm]{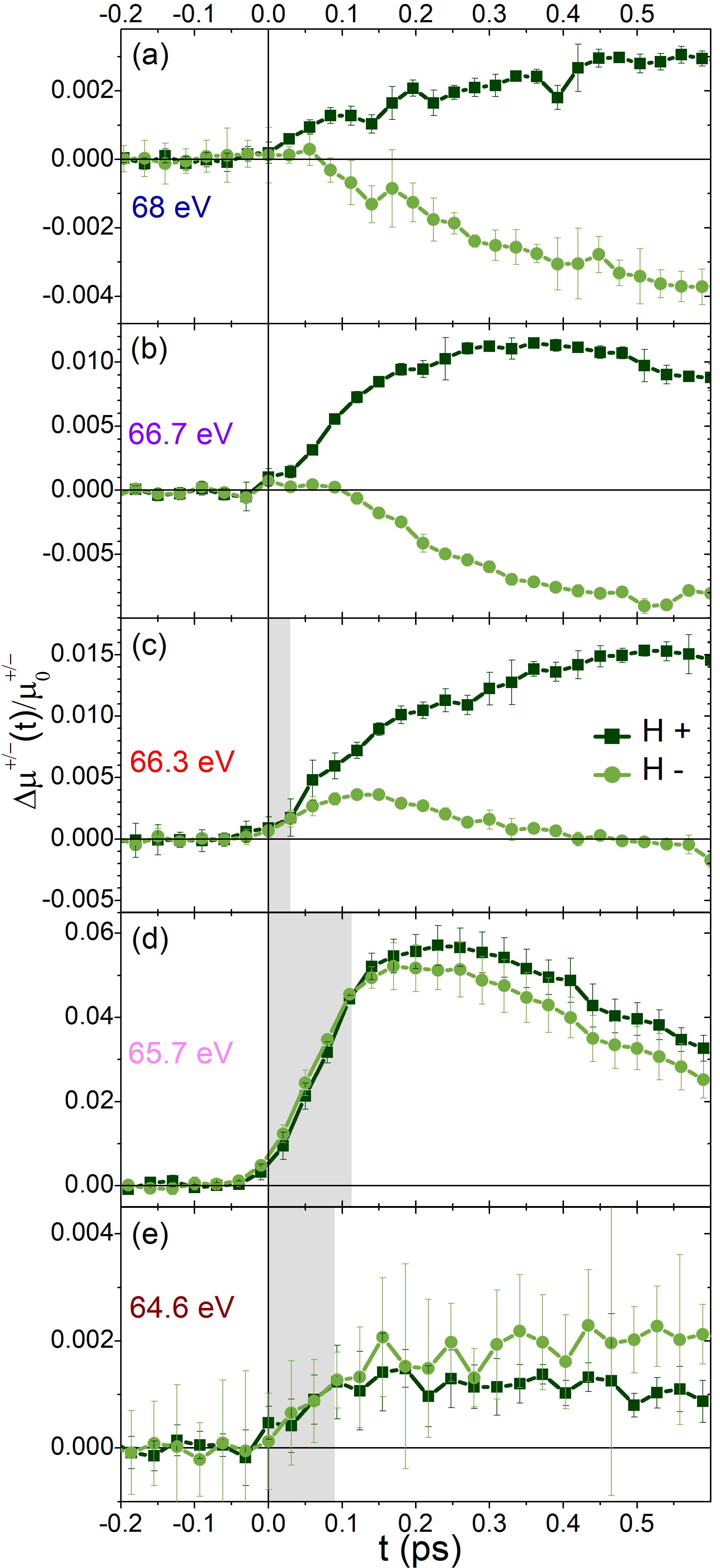}  
    \caption{Relative absorption time traces for both field directions: $H^+$ shown in dark green and $H^-$ shown in light green using XUV probing energies equal to (\textbf{a}) 68 eV. (\textbf{b}) 66.7 eV. (\textbf{c}) 66.3 eV.(\textbf{d}) 65.7 eV. and (\textbf{e}) 64.6 eV.}
    \label{Fig2}
\end{figure}

Figure \ref{Fig2} shows the relative absorption for both field directions, $\frac{\Delta\mu^{\pm}(t)}{\mu^{\pm}_0}$ obtained using 5 different XUV wavelengths. 
For all these probing energies, we observe an increase of the relative absorption for majority electrons ($H^+$, dark green), while the relative minority electron absorption increases for 64.6, 65.7 and 66.3 eV and decreases for 66.7 and 68 eV.
Depending on the probing energy, the differences between the two applied field directions, i.e., between dark and light green curves, are more or less pronounced and transient \textit{relative} absorption changes do not systematically start at $t = 0$ (highlighted grey regions). At first glance, this is in line with the recent observation of a delay between ultrafast electronic and magnetic dynamics \cite{rosner2020simultaneous,yao2020distinct}. Note however that for $E = 65.7$\,eV, a closer look at the first 100\,fs reveals a crossing between $H^+$ and $H^-$ curves. As we will explain in the discussion section, we believe that this is the signature of a spectrum shift.

\begin{figure}[H]
    \centering
    \includegraphics[width=15 cm]{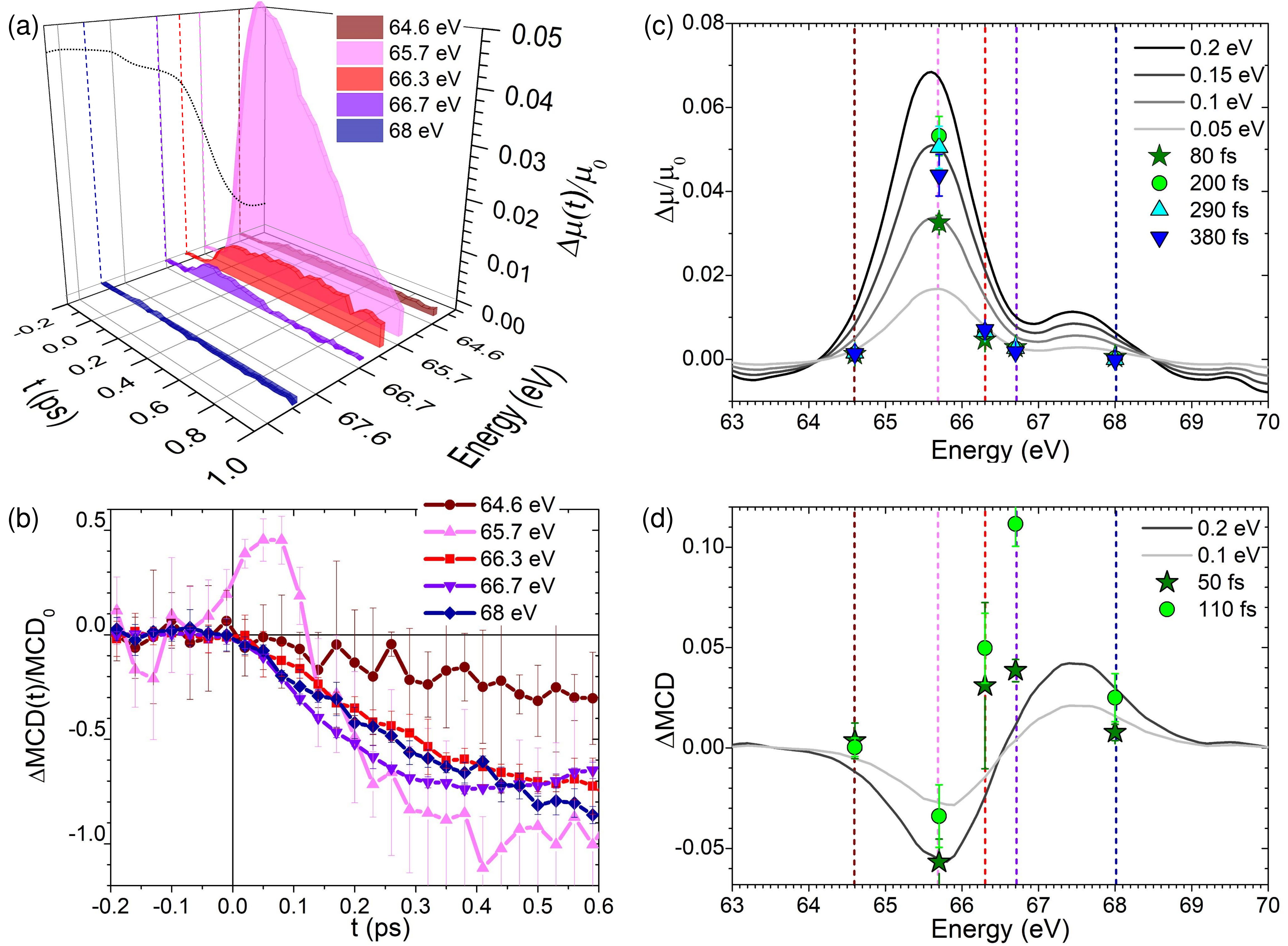}  
    \caption{Summary of time- and energy-dependent absorption and MCD data: (\textbf{a}) Relative absorption time traces as a function of probing photon energies and time delay. (\textbf{b}) Relative MCD, i.e., relative magnetization time traces for the five probing energies. (\textbf{c}) Calculated relative absorption resulting from a shift of the absorption spectra to lower energies (dark and grey lines). The experimental symbols extracted from (\textbf{a}) for different delays are compatible with a maximum shift of -0.15\,eV. (\textbf{d}) Calculated relative magnetization considering a similar MCD spectrum red shift (dark and grey lines). The experimental data, plotted for two different delays, are only in agreement with the prediction resulting from a MCD spectrum shift on very short time scales (<80 fs).}
    \label{Fig3}
\end{figure}

In order to compare the results obtained for different probing energies and further quantify the electronic and magnetic dynamics, we have plotted the relative absorption and the relative MCD in Figure \ref{Fig3} (a) and (b), respectively. Figure \ref{Fig3} (a) shows $\frac{\Delta\mu(t)}{\mu_0}$ (vertical axis) as a function of the delay (left axis) and  probing energies (right axis). As a reminder, the static absorption curve is shown in the background (dotted black line) with the corresponding probing energies (dotted colored line). Figure \ref{Fig3} (b) shows $\frac{\Delta MCD(t)}{MCD_0}$ which is directly proportional to the relative magnetization changes as a function of the time delay for all XUV energies.

Several conclusions can be drawn from these graphs: (i) The relative absorption dynamics depend strongly on the probing energy. In fact, the maximum of the relative transient absorption change observed at 65.7 eV is one order of magnitude stronger than for all other energies scrutinized in this experiment. (ii) The relative magnetization curves display more subtle differences. The characteristic demagnetization times and rates show a weak dependence on the probing energy. As already highlighted, we see that for 66.3 and 64.6 eV, the onset of the demagnetization process is delayed with respect to $t = 0$. For a probing energy of 65.7 eV, we see a surprising increase and subsequent decrease of the magnetization in less than 100 fs. Note that this behavior is seen independently of the helicity of the incoming XUV pulses and has been observed in other Ni samples probed during the same experiment (data not shown).

%%%%%%%%%%%%%%%%%%%%%%%%%%%%%%%%%%%%%%%%%%
\section{Discussion}

When a metallic thin film is excited by an ultrashort optical pulse, the electronic population close to the Fermi level is modified, causing a change of the x-ray absorption spectra on femtosecond timescales. This has first been measured at the Ni $L_3$-edge by Stamm \textit{et al.} \cite{stamm2007femtosecond}, who suggested that optical pumping of the sample can induce a red shift of the absorption edge during the first hundreds of femtoseconds. The effect was theoretically underpinned by Carva \textit{et al.} \cite{carva2009influence}, who explained the observed phenomenon as resulting from an optical excitation-induced electron depletion close to the Fermi energy. Furthermore, their calculations suggested that, in addition to the pure absorption spectra, the MCD should also be affected on sub-picosecond timescales. While subsequent experiments remained limited to the $L$-edges of transition metals \cite{boeglin2010distinguishing}, very recent work by Yao \textit{et al.} \cite{yao2020distinct} and Rösner \textit{et al.} \cite{rosner2020simultaneous} unveiled that similar effects can also be observed at the $M$-edges of Fe, Co and Ni. Furthermore, both studies demonstrated that, depending on the probing energy, a delay between the magnetic and pure charge dynamics (up to 100 fs) can exist. As detailed in Yao's study \cite{yao2020distinct} this might hint at a pump-induced red shift of the MCD spectra. 

To test this hypothesis, we have calculated $\frac{\Delta\mu}{\mu_0}$ and $\Delta MCD$ (using the static calculation shown in Figure \ref{Fig1} (b)), for different red shift values $\Delta E$ ranging from 0.05 eV up to 0.2 eV (see Figure \ref{Fig3} (c) and (d)). On top of these computed curves (black to grey lines), we plot our experimental data using color symbols for different time delays. When considering relative absorption changes, our experimental results are in reasonable agreement with the red shift predictions (Figure \ref{Fig3} (c)). The maximum change of absorption is obtained using $E = 65.7$\,eV after 200\,fs and corresponds to $\Delta E_{\mathrm{max}}$ = -0.15\,eV with respect to the static spectrum. Note that this is in good agreement with the aforementioned $L_3$-edge measurements of Stamm \textit{et al.}, who reported a XAS shift of -0.13\,eV after 200\,fs \cite{stamm2007femtosecond}. All other probe energies give rise to less dramatic changes, in qualitative agreement with our calculations. However, from a quantitative perspective, the observed magnitudes of $\frac{\Delta\mu}{\mu_0}$ show slight deviations from the simple rigid edge shift which hints at more subtle transient modifications of the absorption spectrum \cite{yao2020distinct}.

A comparison of the calculated $\Delta MCD$ with our experimental data, Figure \ref{Fig3} (d), also provides support for a shift of the MCD. Indeed, we can interpret the observed increase of the magnetization shortly after $t = 0$ ($E = 65.7$\,eV, magenta data in panel (b)), as well as the small delays between $t = 0$ and the onset of the demagnetization process ($ E = 64.6$\,eV and, to a lesser extent $ E = 66.3$\,eV for very short times) as resulting from a non-equilibrium electron redistribution around the Fermi level (dark green stars in panel (d)). This increase of the magnetization competes with the usual ultrafast demagnetization, and is visible only briefly after $t = 0$. Note that this explanation challenges a recent study where an increase of the Ni magnetization on early time scales has been attributed to optically induced spin transfer effect \cite{hofherr2017speed}. For $t >$ 100\,fs the magnetization quenching dominates, irrespective of the energy. For probing energies where the electronic redistribution and ultrafast demagnetization channels do not compete (66.7 end 68 eV), we do not find any lag between the onset of the electronic and magnetic dynamics (Figure \ref{Fig2} (a) and (b)).

\begin{figure}[H]
    \centering
    \includegraphics[width=7 cm]{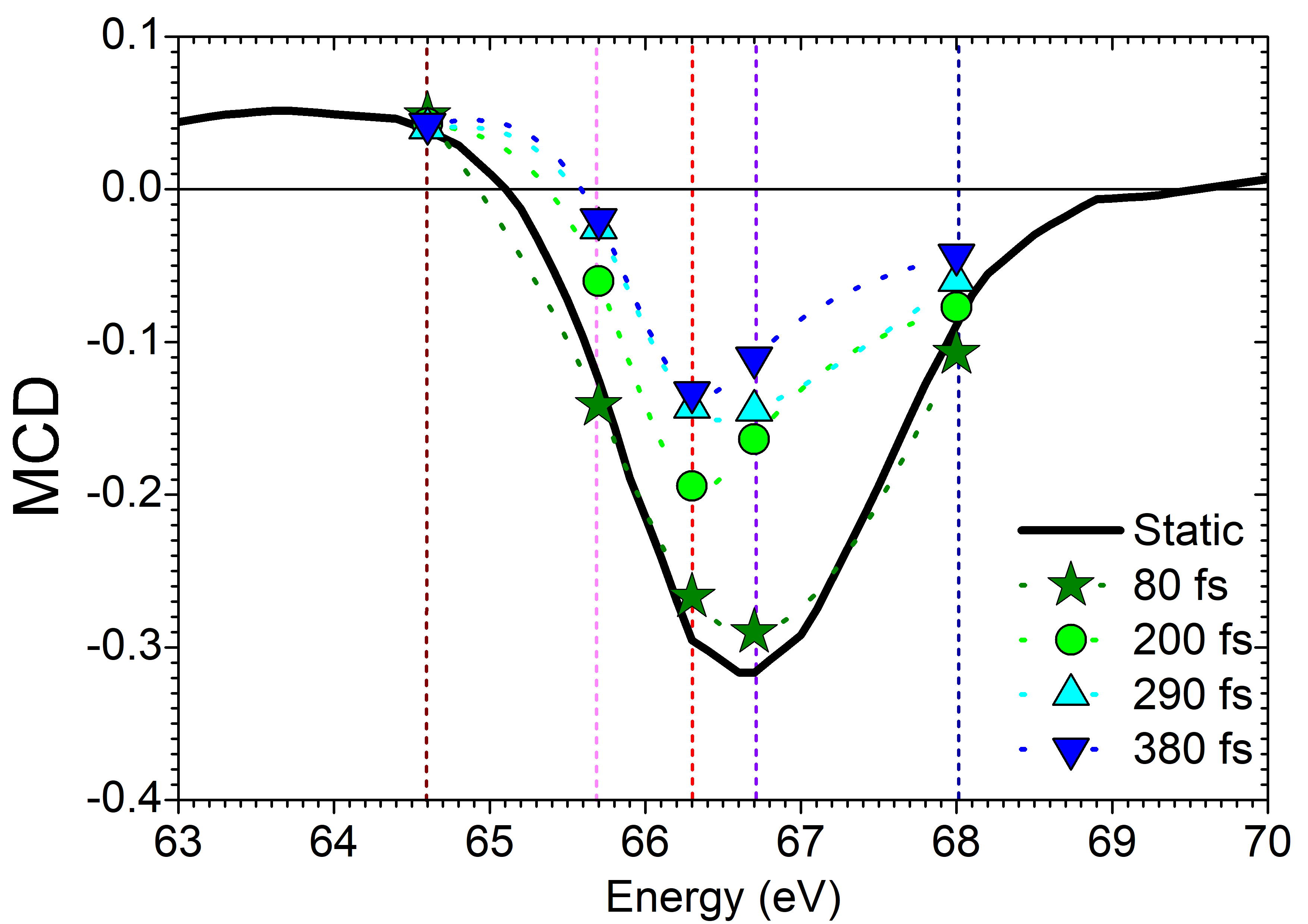}  
    \caption{Calculated static MCD spectra (black line) and measured MCD spectra for different time delays (colored symbols).}
    \label{Fig4}
\end{figure}

The red shift of the MCD should also explicitly be observed when plotting the $MCD$ as a function of the photon energy for different delays. As shown in Figure \ref{Fig4}, we indeed find a slight shift of the static $MCD$ spectrum (black curve) towards smaller energies for $t = 80$\,fs (green stars). On longer timescales however, the MCD magnitude decreases and we observe a distortion of the spectra, especially for higher energies. We emphasize that these results are not at odds with data presented recently by Stamm \textit{et al.} \cite{stamm2020x-ray}. In fact, their measurements on Ni are performed 0.7 ps after the excitation. As shown in the present work, this is well beyond the typical time where $MCD$ shifts can be observed.   

On longer timescales, as already mentioned, clear differences between the different demagnetization curves become apparent (Figure \ref{Fig3} (b)). Despite the large experimental errors (linked to the use of different CCDs for the measurement of the incoming XUV intensity), which impede a thorough quantitative analysis, it is obvious that the demagnetization observed for $E = 64.6$\,eV, is significantly slower than for all other probing energies. This is in agreement with work of Gort \textit{et al.} \cite{gort2018early}, who conclude that the characteristic demagnetization times depend on the binding energy of the probed electrons with respect to the Fermi level. As already suggested by Carva \cite{carva2009influence}, "correctly" probing the magnetization might thus require an integration over the entire MCD spectrum (as intrinsically performed using broadband sources), and must be critically discussed in femtomagnetism experiments, where single, well-defined probing wavelengths are used \cite{carva2009influence}. This could explain why several studies, despite using comparable experimental conditions, reported rather different characteristic demagnetization times \cite{lopez2013role, fan2019timeresolving, hennes2020laserinduced} and significant differences concerning the onset of demagnetization in material systems composed of multiple elements \cite{mathias2012probing, radu2015ultrafast, eschenlohr2017spin}.

%%%%%%%%%%%%%%%%%%%%%%%%%%%%%%%%%%%%%%%%%%
\section{Conclusions}

Using a recently developed x-ray streaking technique \cite{jal2019single, rosner2020simultaneous}, we have probed the photo-absorption cross section of a Ni thin film at the Ni $M_{2,3}$-edges using circularly polarized XUV pulses for five probing energies and both helicities and directions of the applied magnetic field. This allowed us to measure the time dependent absorption and magnetic circular dichroism (MCD) time traces, and to retrieve information about transient changes occurring in the electronic and spin sub-systems close to the Fermi level. Our results show that charge and magnetic dynamics both depend on the probing XUV wavelength. To a first approximation, a red shift of the absorption and MCD spectra on very short timescales (<100-200\,fs) can rationalize our data, in agreement with recent findings \cite{yao2020distinct}. However, our analysis also hints at more complex modifications of those spectra and further work will be required to obtain a truly quantitative picture of femtosecond charge and spin dynamics in magnetic thin films excited by ultrashort laser pulses.

%%%%%%%%%%%%%%%%%%%%%%%%%%%%%%%%%%%%%%%%%%
\vspace{6pt} 

%%%%%%%%%%%%%%%%%%%%%%%%%%%%%%%%%%%%%%%%%%
\authorcontributions{B.R., J.L., C.D., F.C., R.J, B.V., and E.J. conceived the experiments, B.R., F.D. V.A.G, M.L., B.W. and C.D. fabricated the zone plate optics, B.R., R.D., M.L., J.L., A.M., D.N., I.P.N, I.L.Q, E.P., T.S., M.Z., C.D., F.C., B.V., and E.J. performed the experiment at FERMI. Mi.H. and G.M. grew the sample. M.H., V.C., A.K., J.L., F.C., G.S.C, B.V. and E.J. interpreted and discussed the results. M.H. and E.J. prepared the figures and wrote the manuscript. All authors have read and agreed to the published version of the manuscript.}

%%%%%%%%%%%%%%%%%%%%%%%%%%%%%%%%%%%%%%%%%%
\funding{This work was funded within the EU-H2020 Research and Innovation Programme, No. 654360 NFFA-Europe (BR). The authors are grateful for the financial support received from the CNRS-MOMENTUM, the SNSF project (No. 200021-160186), the UMAMI ANR-15-CE24–0009, and the CNRS-PICS programs.}

%%%%%%%%%%%%%%%%%%%%%%%%%%%%%%%%%%%%%%%%%%
\acknowledgments{We acknowledge the great support of the technical team of FERMI, at Trieste. We also thank Clemens von Korff Schmising for stimulating discussions.}

%%%%%%%%%%%%%%%%%%%%%%%%%%%%%%%%%%%%%%%%%%
\conflictsofinterest{The authors declare no conflict of interest. The funders had no role in the design of the study; in the collection, analysis, or interpretation of data; in the writing of the manuscript, or in the decision to publish the results.} 

%%%%%%%%%%%%%%%%%%%%%%%%%%%%%%%%%%%%%%%%%%
%% optional
\abbreviations{The following abbreviations are used in this manuscript:\\

\noindent 
\begin{tabular}{@{}ll}
XUV & Extrem Ultra-Violet \\
MCD & Magnetic Circular Dichroism \\
FEL & Free Electron Laser \\
HHG & High Harmonic Generation \\
CCD & Charge Coupled Device \\
I & Intensity \\
H & Applied Magnetic Field \\
T & Transmission \\
$\mu$ & Absorption
\end{tabular}}


%=====================================
% References, variant B: external bibliography
%=====================================
\externalbibliography{yes}
\bibliography{biblioZP_1colorNi.bib}

%%%%%%%%%%%%%%%%%%%%%%%%%%%%%%%%%%%%%%%%%%
%% optional
\sampleavailability{Sample is available from the authors.}

%% for journal Sci
%\reviewreports{\\
%Reviewer 1 comments and authors’ response\\
%Reviewer 2 comments and authors’ response\\
%Reviewer 3 comments and authors’ response
%}

%%%%%%%%%%%%%%%%%%%%%%%%%%%%%%%%%%%%%%%%%%
\end{document}